\documentclass[twocolumn,aps,prc,superscriptaddress,showpacs,floatfix]{revtex4}
\usepackage{amssymb}
\usepackage{amsmath}
\usepackage{graphicx}

\begin{document}

\title{$\phi $ and $\Omega $ production in relativistic heavy-ion collisions
in a dynamical quark coalescence model}
\author{Lie-Wen Chen}
\affiliation{Institute of Theoretical Physics, Shanghai Jiao Tong University, Shanghai
200240, China}
\affiliation{Center of Theoretical Nuclear Physics, National Laboratory of Heavy Ion
Accelerator, Lanzhou 730000, China}
\author{Che Ming Ko}
\affiliation{Cyclotron Institute and Physics Department, Texas A\&M University, College
Station, Texas 77843-3366}
\date{\today }

\begin{abstract}
Based on the phase-space information obtained from a multi-phase transport
model within the string melting scenario for strange and antistrange quarks,
we study the yields and transverse momentum spectra of $\phi $ mesons and $%
\Omega $ ($\Omega ^{-}+\bar{\Omega}^{+}$) baryons as well as their
anisotropic flows in Au+Au collisions at RHIC using a dynamical quark
coalescence model that includes the effect due to quark phase-space
distributions inside hadrons. With current quark masses and fixing the $\phi 
$ and $\Omega $ radii from fitting measured yields, we first study the ratio
of the yield of $\Omega $ baryons to that of $\phi $ mesons as well as their
elliptic and fourth-order flows as functions of their transverse momentum.
How the elliptic and fourth-order flows of $\phi $ mesons and $\Omega $
baryons are related to those of strange and antistrange quarks is then
examined. The dependence of above results on $\phi $ and $\Omega $ radii as
well as on the strange quark mass is also studied.
\end{abstract}

\pacs{12.38.Mh, 5.75.Ld, 25.75.-q, 24.10.Lx}
\maketitle

\section{Introduction}

\label{introduction}

Recently, there is a lot of interest in using the quark coalescence or
recombination model to understand the experimental data from heavy-ion
collisions at the Relativistic Heavy Ion Collider (RHIC). As shown in Refs. 
\cite{greco,hwa,fries,molnar03}, the quark coalescence model can explain
successfully the observed anomalously large enhancement of baryon to meson
ratio at intermediate transverse momenta and scaling of the elliptic flow of
identified hadrons according to their valence quark numbers. Most of these
applications of the quark coalescence model were based on a simple
momentum-space coalescence in which only quarks with same momentum can
coalesce into hadrons. Also, the phase-space information of quarks in the
partonic matter produced in relativistic heavy-ion collisions is usually
taken from a schematic fireball model. Although the fireball model was still
used in Ref. \cite{greco}, the effect due to quark momentum spread and
spatial distribution inside hadrons was included in calculating the
probabilities for quarks to coalesce to hadrons. Only in Ref. \cite{molnar05}
was the coalescence probability calculated also with a quark phase-space
information from a dynamical parton cascade model. Results from such a
dynamical quark coalescence model indicate that the phase-space structure of
quarks at freeze out plays an important role in hadron production from the
quark-gluon plasma. These studies were mainly concerned with hadrons which
consist of light quarks and/or heavy charm quarks. In the present study, we
use the dynamical quark coalescence model, that is based on the quark
phase-space information from a multi-phase transport (\textrm{AMPT}) model
within the string melting scenario and includes the quark structure of
hadrons, to study instead the production and anisotropic flow of $\phi $
mesons and $\Omega $ ($\Omega ^{-}+\bar{\Omega}^{+}$) baryons that consist
of strange quarks in Au+Au collisions at RHIC.

The study of $\phi $ meson and $\Omega $ baryon production in relativistic
heavy-ion collisions is a topic of great interest as one of the signatures
for the quark-gluon plasma (QGP) produced in relativistic heavy-ion
collisions is enhanced production of hadrons consisting of strange and/or
antistrange quarks. The enhancement occurs because masses of strange quarks
are comparable to the temperature of the QGP and are thus expected to be
abundantly produced from quark and gluon inelastic scattering once the QGP
is formed in the collisions \cite{rafelski82,koch86}. Since the $\phi $
meson carries hidden strangeness ($s\bar{s}$) and the $\Omega $ baryon
consists of three valence strange quarks ($sss$) or antistrange quarks ($%
\bar{s}\bar{s}\bar{s}$), their production in heavy-ion collisions would also
be enhanced if the QGP is formed.

Furthermore, anisotropic flows of $\phi $ mesons and $\Omega $ baryons in
relativistic heavy-ion collisions are useful for understanding the
collective dynamics of strange quarks in the produced QGP \cite%
{ygma,StarMSBflow05}. It is known that anisotropic flows are sensitive to
the properties of dense matter formed during the early stage of heavy-ion
collisions \cite{Ollit92,Rqmd,Danie98,Zheng99}. This sensitivity not only
exists in the elliptic flow $v_{2}$ \cite%
{tean,kolb,huov,Zhang99,moln1,Lin:2001zk,gyulv2,Voloshin03} but also in the
smaller higher-order anisotropic flows, such as the fourth-order anisotropic
flow $v_{4}$ \cite{Kolb99,Teaney99,Kolb00,Kolb03,STAR03,STAR05,chen04,Kolb04}%
. Moreover, observed hadron elliptic flows in heavy-ion collisions at RHIC
were found to satisfy the valence quark number scaling, i.e., the elliptic
flow per quark is the same at same transverse momentum per quark. As shown
in Refs. \cite{greco,fries,molnar03}, such a scaling of hadron elliptic
flows according to their valence quark numbers can be understood in the
quark recombination/coalescence model. Other scaling relations among hadron
anisotropic flows, such as $v_{4}(p_{T})\sim v_{2}^{2}(p_{T})$, have also
been observed in experimental data \cite{STAR03,STAR05}, and they have been
shown in the quark coalescence model to relate to similar scaling relations
among quark anisotropic flows \cite{chen04,Kolb04}. Because of their small
scattering cross sections with other hadrons \cite%
{shor85,singh86,hecke98,bass99,cheng03,biagi81,muller72}, $\phi $ mesons and 
$\Omega $ baryons are not much affected by rescattering effects in later
hadronic stage of the collision and are thus expected to provide more direct
information on the properties of the QGP and how they are produced during
hadronization \cite{StarMSB130,phiStar,phiPhenix,OMGStar}.

Our study shows that in heavy-ion collisions at RHIC the ratio of the yield
of $\Omega $ baryons to that of $\phi $ mesons at intermediate transverse
momenta is strongly enhanced relative to that at low transverse momenta,
similar to that observed in the data for the proton to pion ratio. Also, the
elliptic flows $v_{2}(p_{T})$ of $\phi $ mesons and $\Omega $ baryons follow
approximately the valence quark number scaling. Their valence quark number
scaled elliptic flows deviate, however, strongly from the elliptic flow of
strange and antistrange quarks. For the forth-order anisotropic flow $%
v_{4}(p_{T})$, it scales with the square of the elliptic flow for both $\phi 
$ mesons and $\Omega $ baryons. We have also studied the dependence of above
results on the radii of $\phi $ meson and $\Omega $ baryon as well as on the
strange quark mass. For physically reasonable radii, both yields of $\phi $
mesons and $\Omega $ baryons increase with increasing radii. Their yields
eventually decrease as the radii become unrealistically large. The scaled
elliptic flows of both $\phi $ mesons and $\Omega $ baryons increase with
increasing radii and approach that of strange and antistrange quarks for
very large radii as in the naive momentum-space quark coalescence model.
Changing strange and antistrange quark masses to their constituent quark
masses affects significantly the yields of $\phi $ mesons and $\Omega $
baryons but not much their anisotropic flows. Our results therefore suggest
that both the dynamical phase-space information of partons at freeze out and
quark structure of hadrons are important for $\phi $ meson and $\Omega $
baryon production in relativistic heavy-ion collisions. We note that our
study based on the dynamical quark coalescence model for $\phi $ meson
production and elliptic flow in relativistic heavy-ion collisions is
different from that of Ref. \cite{ygma}, where $\phi $ mesons were
reconstructed from the invariant mass distributions of kaons and antikaons
from the AMPT model.

The paper is organized as follows. In Section \ref{dynamics}, we briefly
review the AMPT model and present detailed information on the dynamics of
quarks and antiquarks (also called partons) in heavy-ion collisions at RHIC.
In Section \ref{coalescence}, the dynamical quark coalescence model based on
parton phase-space distributions at freeze out is described for $\phi $
meson and $\Omega $ baryon production. This also includes the construction
of the quark Wigner phase-space functions inside the $\phi $ meson and $%
\Omega $ baryon. The dynamical quark coalescence model is then used in
Section \ref{production} to study the yields and transverse momentum spectra
as well as the anisotropic flows of $\phi $ mesons and $\Omega $ baryons in
Au+Au collisions at RHIC. In Section \ref{size}, dependence of above results
on the radii of $\phi $ meson and $\Omega $ baryon is studied. The effect
due to change in the strange quark mass is studied in Section \ref{mass}.
Finally, we conclude with a summary in Section \ref{summary}.

\section{Partonic dynamics in heavy-ion collisions at RHIC}

\label{dynamics}

The dynamical quark coalescence model used in present study requires the
space-time and momentum information of quarks and antiquarks during
hadronization of the produced QGP in relativistic heavy-ion collisions. In
particular, the dynamics of strange and antistrange quarks are needed for
describing the production of $\phi $ mesons and $\Omega $ baryons as well as
their anisotropic flows, and they are taken from the AMPT model within the
string melting scenario \cite{ampt}.

\subsection{The AMPT model}

The \textrm{AMPT} model is a hybrid model that uses minijet partons from
hard processes and strings from soft processes in the Heavy Ion Jet
Interaction Generator (\textrm{HIJING}) model \cite{Wang:1991ht} as the
initial conditions for modeling heavy-ion collisions at ultra-relativistic
energies. Time evolution of resulting minijet partons, which are mostly
gluons, is described by Zhang's parton cascade (\textrm{ZPC}) \cite%
{Zhang:1997ej} model. At present, this model includes only parton-parton
elastic scatterings with in-medium cross sections derived from the
lowest-order Born diagrams and having magnitude and angular distribution
fixed by treating the gluon screening mass as a parameter. After minijet
partons stop interacting, they are combined with their parent strings, as in
the \textrm{HIJING} model with jet quenching, to fragment into hadrons using
the Lund string fragmentation model as implemented in the \textrm{PYTHIA}
program \cite{Sjostrand:1994yb}. The final-state hadronic scatterings are
modeled by a relativistic transport (\textrm{ART}) model \cite{Li:1995pr}.

Since the initial energy density in heavy-ion collisions at \textrm{RHIC} is
much larger than the critical energy density at which the hadronic matter to
quark-gluon plasma transition would occur \cite{zhang,chen06,Kharzeev:2001ph}%
, the \textrm{AMPT} model has been extended by converting initial excited
strings into partons \cite{Lin:2001zk}. In this string melting scenario,
hadrons that would have been produced from string fragmentation are
converted instead to their valence quarks and/or antiquarks. Interactions
among these quarks are again described by the\textrm{\ ZPC} parton cascade
model. Since inelastic scatterings are at present not included, the
resulting partonic matter consists of only quarks and antiquarks from melted
strings. To take into account the effect of stronger scattering among gluons
if they were present, the scattering cross sections between quarks and
antiquarks are taken to be the same as those for gluons. These quarks and
antiquarks are converted to hadrons when they stop scattering with other
partons. For a parton cross section of $10$ mb, which is used in present
study, this criteria for hadronization is not too different from the one
based on the condition that the local energy density of hadronizing partons
is at the critical energy density given by lattice QCD calculations. The
transition from the partonic matter to the hadronic matter is achieved using
a simple coalescence model, which combines two nearest quark and antiquark
into mesons and three nearest quarks or antiquarks into baryons or
anti-baryons that are close to the invariant mass of these partons. Details
of the AMPT model can be found in Ref. \cite{ampt}.

In the present study, we use the AMPT model in the string melting scenario
with the default current quark mass of $9.9$ \textrm{MeV} for \textsl{d}
quark, $5.6$ \textrm{MeV} for \textsl{u} quark, and $199$ \textrm{MeV} for 
\textsl{s} quark and a constant parton scattering cross section of $10$ mb.
We note that using parton scattering cross sections of $6$-$10$ \textrm{mb},
the \textrm{AMPT} model with string melting was able to reproduce both the
centrality and transverse momentum (below $2$ \textrm{GeV}$/c$) dependence
of the elliptic flow \cite{Lin:2001zk} and pion interferometry \cite%
{LinHBT02} measured in Au+Au collisions at $\sqrt{s_{NN}}=130$ \textrm{GeV}
at \textrm{RHIC} \cite{Ackermann:2000tr,STARhbt01} as well as measured $%
p_{T} $ dependence of both $v_{2}$ and $v_{4}$ of mid-rapidity charged
hadrons in the same collision at $\sqrt{s_{NN}}=200$ GeV \cite{chen04}.

\subsection{Spectrum of strange and antistrange quarks}

\begin{figure}[th]
\includegraphics[scale=0.9]{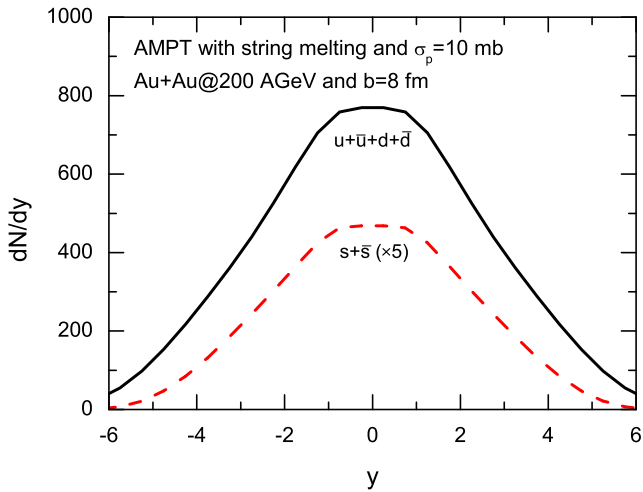}
\caption{{\protect\small (Color online) Rapidity distributions of strange
and antistrange quarks (}$s+\bar{s}${\protect\small ) as well as light
quarks and antiquarks (}$u+\bar{u}+d+\bar{d}${\protect\small ) at freeze out
in Au+Au collisions at }$\protect\sqrt{s_{NN}}=200${\protect\small \ GeV and 
}$b=8${\protect\small \ fm.}}
\label{dNdYq}
\end{figure}

We first show in Fig. \ref{dNdYq} the rapidity distributions of strange
quarks and antiquarks ($s+\bar{s}$) as well as light quarks and antiquarks ($%
u+\bar{u}+d+\bar{d}$) at freeze out, i.e., stop scattering with other
partons, in Au+Au collisions at $\sqrt{s_{NN}}=200$ GeV and $b=8$ fm. It is
seen that the rapidity density $dN/dy$ at mid-rapidity is about $94$ for
strange quarks and antiquarks and about $770$ for light quarks and
antiquarks. Since inelastic partonic scatterings are absent in the present
model, all these quarks and antiquarks are from melted strings during
initial stage of the collisions.

\begin{figure}[th]
\includegraphics[scale=0.9]{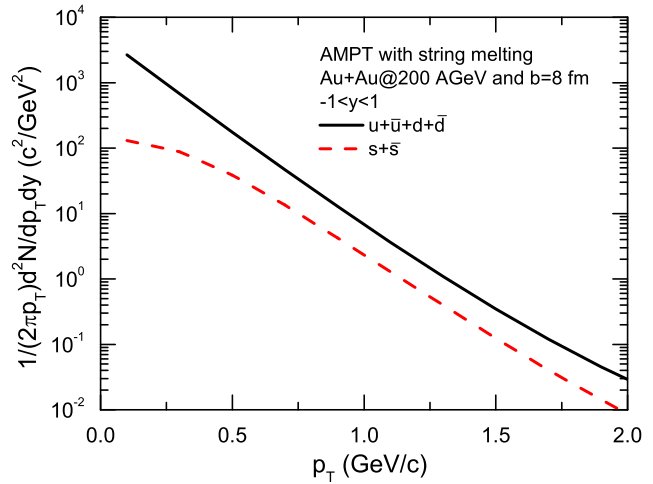}
\caption{{\protect\small (Color online) Transverse momentum distributions of
mid-rapidity strange and antistrange quarks (}$s+\bar{s}${\protect\small )
as well as light quarks and antiquarks (}$u+\bar{u}+d+\bar{d}$%
{\protect\small ) at freeze out in Au+Au collisions at }$\protect\sqrt{s_{NN}%
}=200${\protect\small \ GeV and }$b=8${\protect\small \ fm..}}
\label{dNdPTq}
\end{figure}

The transverse momentum distributions of midrapidity strange and antistrange
quarks as well as light quarks and antiquarks at freeze out in the same
collision are shown in Fig. \ref{dNdPTq}. It clearly shows a quark mass
effect as strange and antistrange quarks display a stiffer transverse
momentum distribution at low $p_{T}$ compared to that of light quarks and
antiquarks. Fitting the transverse momentum spectrum by a Boltzmann
distribution, we find that the effective temperature of the partonic matter
at freeze out is about $161$ MeV, which is consistent with the predicted
critical temperature $T_{c}\approx 150$-$180$ MeV from the lattice QCD for
the QGP to hadronic matter phase transition \cite{karsch02,fodor03}.

\subsection{Space-time structure of strange and antistrange quarks at freeze
out}

\begin{figure}[th]
\includegraphics[scale=1.3]{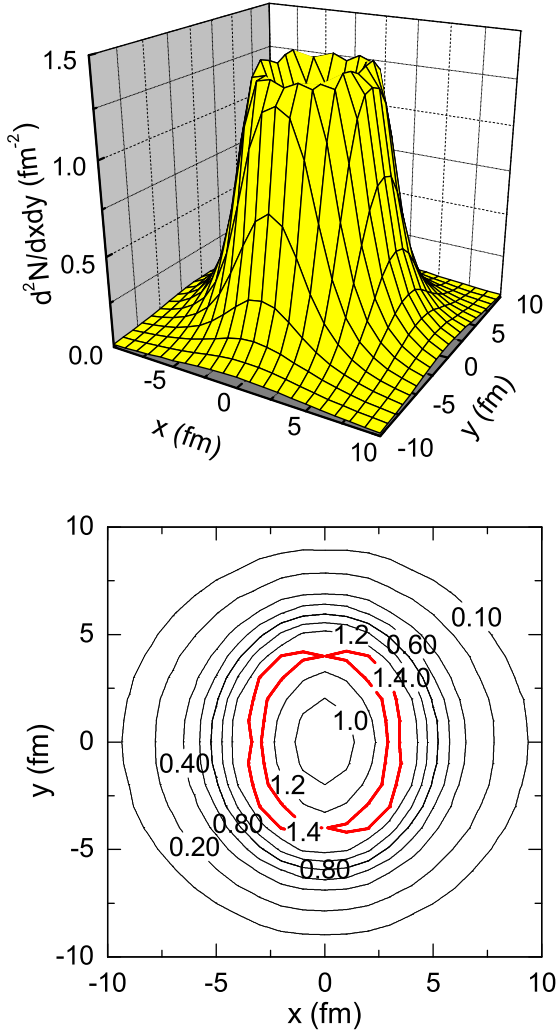}
\caption{{\protect\small (Color online) Number density distribution in the
transverse plane for mid-rapidity strange and antistrange quarks at freeze
out in Au+Au collisions at }$\protect\sqrt{s_{NN}}=200${\protect\small \ GeV
and }$b=8${\protect\small \ fm (upper panel) and corresponding contour plot
(lower panel).}}
\label{dNdxdySSb}
\end{figure}

The spatial distribution of mid-rapidity strange and antistrange quarks at
freeze out in the above collision is shown in Fig. \ref{dNdxdySSb} with the
upper panel displaying their number density distribution in the transverse
plane and the lower panel giving corresponding contour plot. Here, the
transverse plain refers to the $x$-$y$ plane with the $x$-axis pointing to
the direction of the impact parameter and the $y$-axis perpendicular to the $%
x$-axis as well as the beam direction. It is seen that the number density of
strange and antistrange quarks at freeze out peaks approximately at a circle
with radius of about $4$ fm in the transverse plane, implying that these
partons mainly freeze out from the surface of an expanding fireball where
the number density is roughly between $1.2$ fm$^{-2} $ and $1.4$ fm$^{-2}$
in the transverse plane.

\begin{figure}[th]
\includegraphics[scale=0.85]{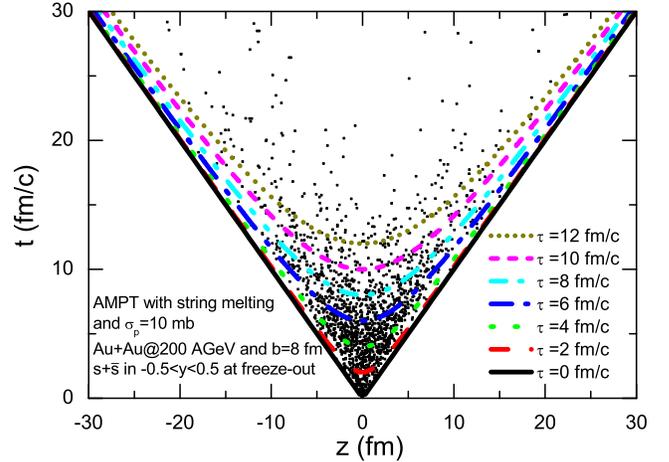}
\caption{{\protect\small (Color online) Space-time structure of mid-rapidity
strange and antistrange quarks at freeze out in Au+Au collisions at }$%
\protect\sqrt{s_{NN}}=200${\protect\small \ GeV and } $b=8${\protect\small \
fm.}}
\label{ztSSb}
\end{figure}

\begin{figure}[th]
\includegraphics[scale=0.95]{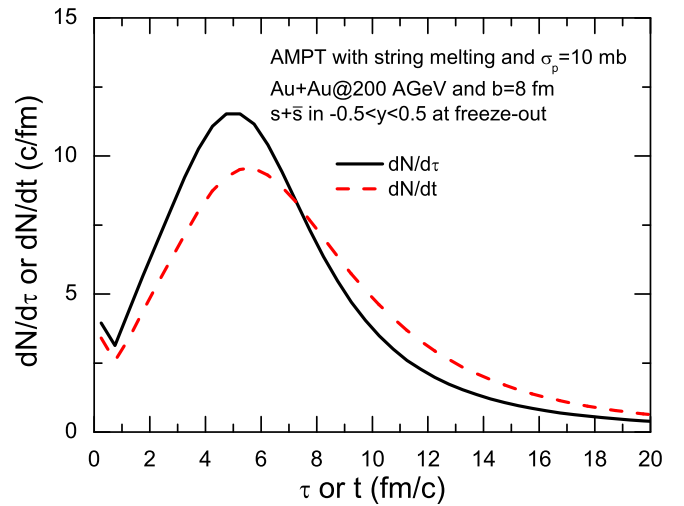}
\caption{{\protect\small (Color online) Freeze out rate of mid-rapidity
strange and antistrange quarks as a function of proper time }$\protect\tau $%
{\protect\small \ as well as the time }$t${\protect\small \ in the
nucleus-nucleus center-of-mass system.}}
\label{dNdTAUtSSb}
\end{figure}

The space-time structure of $z$-$t$ correlations for mid-rapidity strange
and antistrange quarks at freeze out in this collision is shown in Fig. \ref%
{ztSSb}. Instead of a sudden freeze out, these partons are seen to freeze
out continuously in the proper time $\tau $, with most between $\tau =4$ and 
$6$ fm/c. This can be more clearly seen from their freeze out rate as a
function of proper time $\tau $ or the time $t$ in the nucleus-nucleus
center-of-mass system as shown in Fig. \ref{dNdTAUtSSb}. It shows that the
freeze out rate of midrapidity strange and antistrange quarks peaks at $\tau
=5$ fm/c or $t=5.5$ fm/c.

\begin{figure}[th]
\includegraphics[scale=0.85]{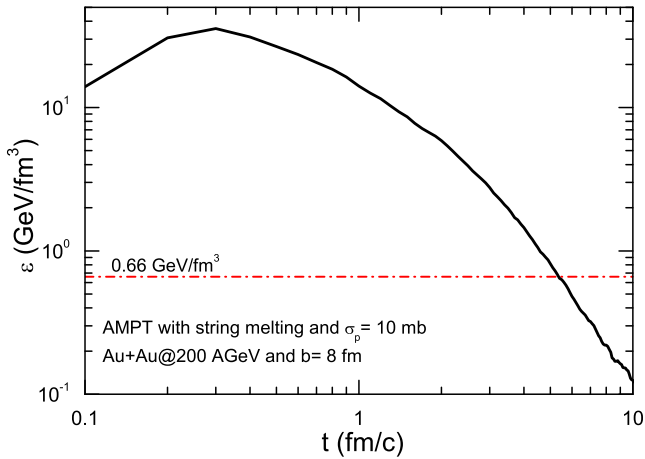}
\caption{{\protect\small (Color online) Time evolution of the energy density
in the central cell of partonic matter in Au+Au collisions at }$\protect%
\sqrt{s_{NN}}=200${\protect\small \ GeV and }$b=8${\protect\small \ fm.}}
\label{eDen}
\end{figure}

The AMPT model also provides information on the time evolution of parton
energy density. This is shown in Fig. \ref{eDen} for the central cell of the
partonic matter in above collision. As in Refs. \cite{zhang,chen06}, the
central cell is taken to have a transverse radius of $1$ fm and a
longitudinal dimension of $5\%$ of the time $t$ after the two nuclei have
fully overlapped in the longitudinal direction. It is seen that the central
energy density is about $0.6$ GeV/fm$^{3}$ at $t=5.5$ fm/c when most of
partons freeze out. This energy density is consistent with the critical
energy density of about $0.66$ GeV/fm$^{3}$ predicted by the lattice QCD for
the quark-gluon plasma to hadronic matter transition \cite{karsch02,fodor03}.

\subsection{Anisotropic flow of strange and antistrange quarks}

Anisotropy in the transverse momentum distribution of particles in
non-central heavy-ion collisions is generated by the pressure anisotropy in
the initial compressed matter \cite{Barrette94, Appel98} and thus depends on
the geometry and energy density as well as the properties of produced matter
during the early stage of these collisions. For $\phi $ mesons and $\Omega $
baryons, their anisotropic flows are essentially determined by those of
strange and antistrange quarks at freeze out and their hadronization
mechanism. As discussed previously, the effect due to later hadronic
scattering is small because the interactions of $\phi $ mesons and $\Omega $
baryons in the hadronic matter are relatively weak and also the pressure
becomes more isotropic during the hadronic stage of the collisions.

From the momentum distribution of strange and antistrange quarks at freeze
out, their elliptic and fourth-order anisotropic flows can be evaluated
according to following single-particle averages: 
\begin{eqnarray}
v_{2}(p_{T}) &=&\left\langle \frac{p_{x}^{2}-p_{y}^{2}}{p_{T}^{2}}%
\right\rangle \\
v_{4}(p_{T}) &=&\left\langle \frac{p_{x}^{4}-6p_{x}^{2}p_{y}^{2}+p_{y}^{4}}{%
p_{T}^{4}}\right\rangle
\end{eqnarray}%
where $p_{x}$ and $p_{y}$ are, respectively, projections of their momentum
in and perpendicular to the reaction plane, defined by the beam and impact
parameter axes, and $p_{T}=(p_{x}^{2}+p_{y}^{2})^{1/2}$ is the transverse
momentum.

\begin{figure}[th]
\includegraphics[scale=0.9]{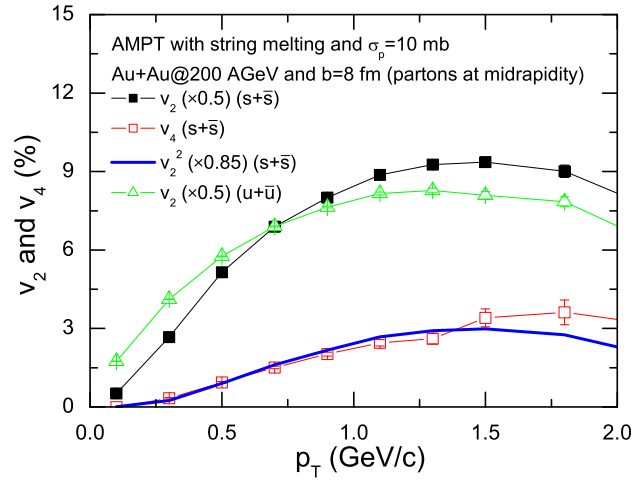}
\caption{{\protect\small (Color online) Transverse momentum} {\protect\small %
dependence of elliptic flow }$v_{2}$ {\protect\small (solid squares) and
forth-order anisotropic flow} $v_{4}$ {\protect\small (open squares) of
mid-rapidity strange and antistrange quarks (}$s+\bar{s}${\protect\small )
at freeze out in Au+Au collisions at }$\protect\sqrt{s_{NN}}=200$%
{\protect\small \ GeV and }$b=8${\protect\small \ fm. The thick solid line
represents }$0.85v_{2}^{2}$ {\protect\small of strange and antistrange
quarks, and open triangles are} $v_{2}${\protect\small \ of mid-rapidity
light up and anti-up quarks (}$u+\bar{u}${\protect\small ).}}
\label{v24PTssBuuB}
\end{figure}

In Fig. \ref{v24PTssBuuB}, we show the $p_{T}$ dependence of $v_{2}$ and $%
v_{4}$ of mid-rapidity strange and antistrange quarks ($s+\bar{s}$) at
freeze out in Au+Au collisions at $\sqrt{s_{NN}}=200$ GeV and $b=8$ fm. It
is seen that strange and antistrange quarks exhibit not only a strong $v_{2}$
but also a non-negligible $v_{4}$ that approximately scales with $v_{2}^{2}$
with a scaling coefficient of 0.85, i.e, $v_{4}\sim 0.85v_{2}^{2}$ as shown
by the thick solid line in Fig. \ref{v24PTssBuuB}. Compared to the $v_{2}$
of mid-rapidity light up and anti-up quarks ($u+\bar{u}$), shown by
triangles in Fig. \ref{v24PTssBuuB}, $v_{2}$ of heavier strange and
antistrange quarks has a smaller value at low $p_{T}$ but a larger value at
high $p_{T}$, similar to the mass ordering of hadron elliptic flows in the
hydrodynamic model which assumes that the produced matter is in local
thermal equilibrium and thus develops a large collective radial flow.
However, instead of a continuing increase of $v_{2}$ with respect to $p_{T}$
as in the hydrodynamic model, $v_{2}$ of partons in the transport model
saturates at a maximum value when their $p_{T}$ becomes large, indicating
that high momentum partons do not reach thermal equilibrium with the bulk of
the partonic matter.

\section{A dynamical quark coalescence model for $\protect\phi $ and $\Omega 
$ production}

\label{coalescence}

In the coalescence model, the probability for forming a bound cluster from a
many-particle system is determined by the overlap of the wave functions of
coalescing particles with the internal wave function of the cluster. Its
validity is based on the assumption that coalescing particles are
statistically independent and the binding energy of formed cluster as well
as the quantum dynamics of the coalescing process play only minor roles \cite%
{Mattie95}. In the present study, we assume that correlations among partons
at freeze out are weak and binding energies of formed hadrons can be
neglected. Furthermore, the coalescence model is considered as a
perturbative approach, valid only if the number of partons coalesced into
hadrons is small compared with the total number of partons in the system.
This condition is satisfied for $\phi $ mesons and $\Omega $ baryons
produced in relativistic heavy-ion collisions as their numbers measured in
experiments are indeed significantly smaller than that of kaons and thus
that of strange and antistrange quarks.

\subsection{The dynamical coalescence model}

In the dynamical quark coalescence model, the probability for producing a
hadron from partons in the QGP is given by the overlap of parton phase-space
distributions at freeze out with the parton Wigner phase-space function
inside the hadron. For a QGP containing $A$ partons, the momentum
distribution of a $M$-parton hadron can be expressed as \cite%
{Mattie95,chen03}%
\begin{eqnarray}
&&\frac{dN_{M}}{d^{3}K}=G\left( 
\begin{array}{c}
A \\ 
M%
\end{array}%
\right) \frac{1}{A^{M}}\int \prod_{i=1}^{M}f_{i}(\mathbf{r}_{i},\mathbf{k}%
_{i})  \notag \\
&&\times \rho ^{W}(\mathbf{r}_{i_{1}},\mathbf{k}_{i_{1}}\cdots \mathbf{r}%
_{i_{M-1}},\mathbf{k}_{i_{M-1}})  \notag \\
&&\times \delta (\mathbf{K}-(\mathbf{k}_{1}+\cdots +\mathbf{k}_{M}))d\mathbf{%
r}_{1}d\mathbf{k}_{1}\cdots d\mathbf{r}_{M}d\mathbf{k}_{M},  \label{Coal}
\end{eqnarray}%
where $f_{i}$ is the parton phase-space distribution functions at
freeze-out; $\rho ^{W}$ is the $M$-parton Wigner phase-space function inside
the hadron; $\mathbf{r}_{i_{1}},\cdots ,\mathbf{r}_{i_{M-1}}$ and $\mathbf{k}%
_{i_{1}},\cdots ,\mathbf{k}_{i_{M-1}}$ are, respectively, the $M-1$ relative
coordinates and momenta taken at equal time in the $M$-parton rest frame ($%
\mathbf{K=0}$); and $G$ is the statistical factor for the $M$ partons to
form the hadron.

In transport model simulations of heavy-ion collisions, the multiplicity of
a $M$-parton hadron produced from the dynamical quark coalescence model is
then given by \cite{Mattie95,chen03} 
\begin{eqnarray}
N_{M} &=&G\int \underset{i_{1}>i_{2}>...>i_{M}}{\sum }d\mathbf{r}_{i_{1}}d%
\mathbf{k}_{i_{1}}\cdots d\mathbf{r}_{i_{M-1}}d\mathbf{k}_{i_{M-1}}  \notag
\\
&\times &\langle \rho _{i}^{W}(\mathbf{r}_{i_{1}},\mathbf{k}_{i_{1}}\cdots 
\mathbf{r}_{i_{M-1}},\mathbf{k}_{i_{M-1}})\rangle ,  \label{MultCluster}
\end{eqnarray}%
where $\langle \cdots \rangle $ denotes event averaging and the sum runs
over all possible combinations of $M$ partons.

To determine the statistical factor $G$ for the $\phi $ meson, we note that
its quark wave function in the color-spin-isospin space can be expressed as
a linear combination of all possible orthogonal flavor, color, and spin
basis states \cite{greco,chen04TH}. The probability for a strange quark $s$
and an anti-strange quark $\bar{s}$ to form a hadron with quantum numbers
corresponding to a $\phi $ meson with the $z$-component of its spin equal to 
$1$ is simply given by the probability of finding the two $s$ and $\bar{s}$
in any one of these color-spin-isospin basis states, i.e., $1/3^{2}\times
1/2^{2}=1/36$. Including also the possibility of forming a $\phi $ meson
with the $z$-component of its spin equal to $-1$ and $0$ triples the
probability. As a result, the statistical factor $G$ for the $\phi $ meson
is $1/12$. A similar consideration leads to a statistical factor $G=1/54$
for the $\Omega ^{-}$ and $\bar{\Omega}^{+}$ baryons.

\subsection{Parton phase-space distributions}

The $M$-parton phase-space distribution function $\Pi _{i=1}^{M}f(\mathbf{r}%
_{i},\mathbf{k}_{i})$ in Eq. (\ref{Coal}) refers to partons with spatial
coordinates and momenta at equal time in the rest frame of the $M$-parton
cluster. Since parton momenta in the AMPT model are given in nucleus-nucleus
center-of-mass system, a Lorentz transformation is performed to obtain their
momenta in the rest frame of the $M$-parton cluster. Furthermore, partons
freeze out at different times as shown in Figs. \ref{ztSSb} and \ref%
{dNdTAUtSSb}. The same Lorentz transformation is thus used in obtaining
their space-time coordinates in the rest frame of the $M$-parton cluster. To
determine the spatial coordinates of these partons at equal time in their
rest frame, partons in the cluster that freeze out earlier are allowed to
propagate freely, i.e., with constant velocities given by the ratio of their
momenta and energies in the rest frame of the cluster, until the time when
the last parton in the cluster freezes out.

\subsection{Quark Wigner phase-space functions inside $\protect\phi $ and $%
\Omega $}

To determine the quark Wigner phase-space functions inside hadrons requires
knowledge of their quark wave functions. For the $\phi $ meson, we take its
quark wave function to be that of a spherical harmonic oscillator, i.e., 
\begin{equation}
\psi (\mathbf{r}_{1},\mathbf{r}_{2})=1/(\pi \sigma _{\phi }^{2})^{3/4}\exp
(-r^{2}/(2\sigma _{\phi }^{2})),  \label{phiWF}
\end{equation}%
in terms of the relative coordinate $\mathbf{r}=\mathbf{r}_{1}-\mathbf{r}%
_{2} $ and the size parameter $\sigma _{\phi }$. The above normalized wave
function leads to a root mean-square radius $R_{\phi }=\left\langle
r^{2}\right\rangle ^{1/2}=\allowbreak (3/8)^{1/2}\sigma _{\phi }$ for the $%
\phi $ meson.

The quark Wigner phase-space function inside the $\phi $ meson is obtained
from its quark wave function by 
\begin{eqnarray}
&&\rho _{\phi }^{W}(\mathbf{r},\mathbf{k})  \notag \\
&=&\int \psi \left( \mathbf{r}+\frac{\mathbf{R}}{2}\right) \psi ^{\ast
}\left( \mathbf{r}-\frac{\mathbf{R}}{2}\right) \exp (-i\mathbf{k}\cdot 
\mathbf{R)}d^{3}\mathbf{R}  \notag \\
&=&8\exp \left( -\frac{r^{2}}{\sigma _{\phi}^{2}}-\sigma _{\phi}^{2}k^{2}
\right)  \label{WignerPhi}
\end{eqnarray}%
where $\mathbf{k}=(\mathbf{k}_{1}-\mathbf{k}_{2})/2$ is the relative
momentum between $s$ and $\bar s$ quarks.

For $\Omega ^{-}$ and $\bar{\Omega}^{+}$ baryons, their quark wave functions
are taken to be the same and are given by that of a spherical harmonic
oscillator as well \cite{Dover83,Heinz99}, i.e., 
\begin{equation}
\psi (\mathbf{r}_{1},\mathbf{r}_{2},\mathbf{r}_{3})=(3\pi ^{2}\sigma
_{\Omega }^{4})^{-3/4}\exp \left( -\frac{\mathbf{\rho }^{\mathbf{2}}+\mathbf{%
\lambda }^{\mathbf{2}}}{2\sigma _{\Omega }^{2}}\right) ,  \label{omgWF}
\end{equation}%
in terms of the relative coordinates $\mathbf{\rho }$ and $\mathbf{\lambda }%
\ $, and the size parameter $\sigma _{\Omega }$. Here we have used the usual
Jacobi coordinates for a three-particle system \cite{chen03}, i.e.,%
\begin{equation}
\left( 
\begin{array}{c}
\mathbf{R} \\ 
\mathbf{\rho } \\ 
\mathbf{\lambda }%
\end{array}%
\right) =\left( 
\begin{array}{ccc}
\frac{1}{3} & \frac{1}{3} & \frac{1}{3} \\ 
\frac{1}{\sqrt{2}} & -\frac{1}{\sqrt{2}} & 0 \\ 
\frac{1}{\sqrt{6}} & \frac{1}{\sqrt{6}} & -\frac{2}{\sqrt{6}}%
\end{array}%
\right) \left( 
\begin{array}{c}
\mathbf{r}_{1} \\ 
\mathbf{r}_{2} \\ 
\mathbf{r}_{3}%
\end{array}%
\right) ,  \label{Jacobi1}
\end{equation}%
where $\mathbf{R}$ is the center-of-mass coordinate of the three quarks or
antiquarks.

Using $d\mathbf{r}_{1}d\mathbf{r}_{2}d\mathbf{r}_{3}=3^{3/2}d\mathbf{R}d%
\mathbf{\rho }d\mathbf{\lambda }$, it is easy to check that the wave
function given by Eq. (\ref{omgWF}) is normalized to one. From the relation $%
(\mathbf{r}_{1}-\mathbf{R)}^{\mathbf{2}}+(\mathbf{r}_{2} -\mathbf{R)}^{%
\mathbf{2}}+(\mathbf{r}_{3}-\mathbf{R)}^{\mathbf{2}} =\mathbf{\rho }^{%
\mathbf{2}}+\mathbf{\lambda }^{\mathbf{2}}$, the root mean-square radius $%
R_\Omega$ of the $\Omega$ baryon is given by 
\begin{equation}
R_\Omega=\left[\int \frac{\mathbf{\rho }^{\mathbf{2}}+\mathbf{\lambda }^{%
\mathbf{2}}}{3}|\psi (\mathbf{r}_{1},\mathbf{r}_{2},\mathbf{r}%
_{3})|^{2}3^{3/2}d\mathbf{\rho }d\mathbf{\lambda }\right]^{1/2}=\sigma
_\Omega.  \label{rms}
\end{equation}

The quark Wigner phase-space function inside the $\Omega $ baryon is
obtained from its quark wave function via 
\begin{eqnarray}
&&\rho _{\Omega }^{W}(\mathbf{\rho },\mathbf{\lambda },\mathbf{k}_{\mathbf{%
\rho }},\mathbf{k}_{\mathbf{\lambda }})  \notag \\
&=&\int \psi \left( \mathbf{\rho +}\frac{\mathbf{R}_{\mathbf{1}}}{2},\mathbf{%
\lambda +}\frac{\mathbf{R}_{\mathbf{2}}}{2}\right) \psi ^{\ast }\left( 
\mathbf{\rho -}\frac{\mathbf{R}_{\mathbf{1}}}{2},\mathbf{\lambda -}\frac{%
\mathbf{R}_{\mathbf{2}}}{2}\right)  \notag \\
&&\times \exp (-i\mathbf{k}_{\mathbf{\rho }}\cdot \mathbf{R}_{\mathbf{1}%
})\exp (-i\mathbf{k}_{\mathbf{\lambda }}\cdot \mathbf{R}_{\mathbf{2}%
})3^{3/2}d\mathbf{R}_{\mathbf{1}}d\mathbf{R}_{\mathbf{2}}  \notag \\
&=&8^{2}\exp \left( -\frac{\mathbf{\rho }^{\mathbf{2}}+\mathbf{\lambda }^{%
\mathbf{2}}}{\sigma _{\Omega }^{2}}\right) \exp (-(\mathbf{k}_{\mathbf{\rho }%
}^{2}+\mathbf{k}_{\mathbf{\lambda }}^{2})\sigma _{\Omega }^{2}),
\label{WignerOMG}
\end{eqnarray}%
where $\mathbf{k}_{\mathbf{\rho }}$ and $\mathbf{k}_{\mathbf{\lambda }}$ are
relative momenta, which together with the total momentum $\mathbf{K}$ are
defined by \cite{chen03} 
\begin{equation}
\left( 
\begin{array}{c}
\mathbf{K} \\ 
\mathbf{k}_{\mathbf{\rho }} \\ 
\mathbf{k}_{\mathbf{\lambda }}%
\end{array}%
\right) =\left( 
\begin{array}{ccc}
1 & 1 & 1 \\ 
\frac{1}{\sqrt{2}} & -\frac{1}{\sqrt{2}} & 0 \\ 
\frac{1}{\sqrt{6}} & \frac{1}{\sqrt{6}} & -\frac{2}{\sqrt{6}}%
\end{array}%
\right) \left( 
\begin{array}{c}
\mathbf{k}_{1} \\ 
\mathbf{k}_{2} \\ 
\mathbf{k}_{3}%
\end{array}%
\right) ,  \label{Jacobi4}
\end{equation}%
with $\mathbf{k}_{1}$, $\mathbf{k}_{2}$, and $\mathbf{k}_{3}$ being the
momenta of the three quarks.

\subsection{$\protect\phi$ and $\Omega$ size parameters}

The two parameters $\sigma _{\phi }$ and $\sigma _{\Omega }$ in the quark
Wigner phase-space functions inside the $\phi $ meson and $\Omega $ baryon
are related to their root-mean-square radii. Since the latter are not known
empirically, we take them as adjustable parameters and fix them by fitting
measured yields of $\phi $ mesons and $\Omega $ baryons in relativistic
heavy-ion collisions.

For $\phi $ mesons, experimental data on their rapidity density $dN/dy$ at
mid-rapidity and its centrality dependence are available for Au+Au
collisions at $\sqrt{s_{NN}}=200$ GeV \cite{phiStar,phiPhenix}. There is,
however, a significant difference between the STAR and PHENIX data for the $%
\phi $ meson yield. For instance, the $\phi $ meson $dN/dy$ at mid-rapidity
is about $7.7$ in the STAR data while $4.5$ in the PHENIX data for Au+Au
collisions at $\sqrt{s_{NN}}=200$ GeV with a centrality of $0$-$5\%$. Using
the above described dynamical quark coalescence model, we find that the STAR
(PHENIX) data can be approximately reproduced with a $\phi $ meson size
parameter $\sigma _{\phi}\approx 1.06$ ($0.77)$ fm, which gives a reasonable 
$\phi $ meson root-mean-square radius of about $R_{\phi }=0.65$ ($0.47)$ fm.
For the impact parameter $b=8$ fm as mainly considered in present study, the
AMPT model gives the number of participant nucleons N$_{\text{part}}\approx
160$ which corresponds to a centrality of about $31\%$. The resulting $\phi $
meson rapidity density $dN/dy$ at mid-rapidity obtained with the $\phi $
meson radius $R_{\phi }=0.65$ ($0.47$) fm is about $3.3$ ($2.0$), which is
also in good agreement with experimental data \cite{phiStar,phiPhenix}.

For $\Omega $ baryons, there are only preliminary experimental data on their 
$dN/dy$ at mid-rapidity in central collisions of Au+Au at $\sqrt{s_{NN}}=200$
GeV, and its value is about $0.64$ \cite{OMGStar}. Fitting the result from
the dynamical quark coalescence model for the same collision at $b=0$ fm
leads to an $\Omega $ baryon size parameter $\sigma _{\Omega }\approx 1.2$
fm, corresponding to an $\Omega $ baryon root-mean-square radius of about $%
R_{\Omega }\approx 1.2$ fm, which is somewhat large but not unrealistic. For
more peripheral collisions at impact parameter $b=8$ fm, an $\Omega $ radius
of $R_{\Omega }=1.2$ fm leads to a rapidity density $dN/dy$ of about $0.26$
at mid-rapidity.

In the following, we use the above determined $\phi$ and $\Omega$ root
mean-square radii, i.e., $R_{\phi }=0.65$ fm or $0.47$ fm for the $\phi $
meson while $R_{\Omega }=1.2$ fm for the $\Omega $ baryon unless stated
otherwise.

\section{$\protect\phi $ and $\Omega $ production in Au+Au collisions at $%
\protect\sqrt{s_{NN}}=200~\mathrm{GeV}$}

\label{production}

Using the phase-space information and the dynamical coalescence model
described above, we can now study the transverse momentum spectra of $\phi $
mesons and $\Omega $ baryons as well as their anisotropic flows in
relativistic heavy-ion collisions.

\subsection{Transverse momentum spectra of $\protect\phi $ and $\Omega $}

\begin{figure}[th]
\includegraphics[scale=1.0]{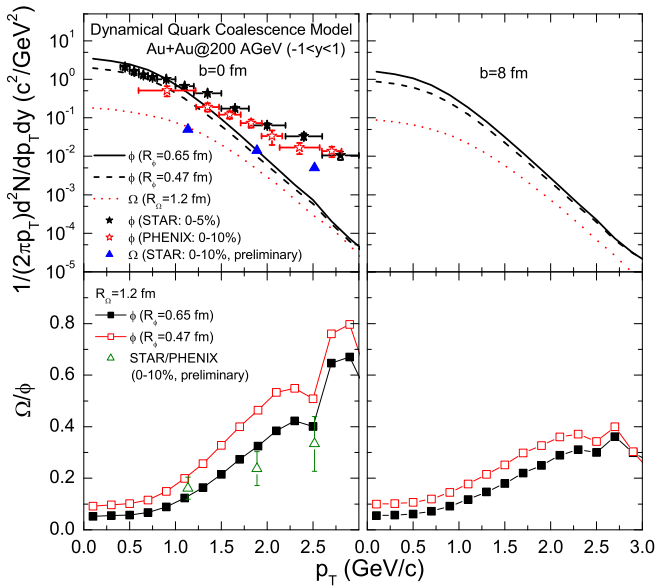}
\caption{{\protect\small (Color online) }{\protect\small Transverse momentum
dependence of midrapidity }$\protect\phi ${\protect\small \ mesons and }$%
\Omega ${\protect\small \ (}$\Omega ^{-}+\bar{\Omega}^{+}${\protect\small )
baryons (upper panels) and their ratio (lower panels) in Au+Au collisions at 
}$\protect\sqrt{s_{NN}}=200${\protect\small \ GeV with }$b=0$ 
{\protect\small \ fm (left panels) and }$b=8${\protect\small \ fm (right
panels). Data for }$\protect\phi ${\protect\small \ mesons in }$0$ 
{\protect\small -}$5\%${\protect\small \ centrality bin\ from the STAR
collaboration \protect\cite{phiStar} (solid stars) and in }$0$%
{\protect\small -}$10\%${\protect\small \ centrality bin from the PHENIX
collaboration \protect\cite{phiPhenix} (open stars) as well as preliminary
data for }$\Omega ${\protect\small \ baryons in }$0${\protect\small -}$10\%$%
{\protect\small \ centrality bin\ from the STAR collaboration \protect\cite%
{OMGStar} (solid triangles) are shown in upper left panel, while those for
the }$\Omega /\protect\phi ${\protect\small \ ratio in }$0${\protect\small -}%
$10\%${\protect\small \ centrality bin\ are shown in lower-left panel.}}
\label{dNdPTratioPhiOmg}
\end{figure}

Fig. \ref{dNdPTratioPhiOmg} shows the transverse momentum spectra of
midrapidity $\phi $ mesons and $\Omega $ baryons as well as their ratio in
Au+Au collisions at $\sqrt{s_{NN}}=200$ GeV with $b=0$ fm (left panels) and $%
b=8$ fm (right panels), and $R_{\phi }=0.65$ fm or $0.47$ fm and $R_{\Omega
}=1.2$ fm. For comparison, we also include in upper left panel of Fig. \ref%
{dNdPTratioPhiOmg} the experimental data for $\phi $ mesons in $0$-$5\%$
centrality bin from the STAR collaboration \cite{phiStar} (solid stars) and
in $0$-$10\%$ centrality bin from the PHENIX collaboration \cite{phiPhenix}
(open stars) as well as the preliminary data for $\Omega $ baryons in $0$-$%
10\%$ centrality bin from the STAR collaboration \cite{OMGStar} (solid
triangles). Also shown in lower left panel by open triangles are data on the 
$\Omega /\phi $\ ratio in $0$-$10\%$ centrality bin, obtained from a linear
interpolation of the $\phi $ meson data in upper left panel.

It is seen from upper left panel of Fig. \ref{dNdPTratioPhiOmg} that the
calculated transverse momentum spectrum of $\Omega $ baryons is stiffer than
that of $\phi $ mesons as expected from their different masses. Both
calculated transverse momentum spectra of $\phi $ mesons and $\Omega $
baryons are, however, significantly softer than the data. With $R_{\phi
}=0.65$ ($0.47$) fm, the mean transverse momentum calculated from the $\phi $
meson transverse momentum spectrum is about $0.72 $ ($0.73$) and $0.68$ ($%
0.70$) GeV/c for $b=0$ fm and $b=8$ fm, respectively, compared with the
experimental value of about $0.85$-$1.1$ GeV/c \cite{phiStar,phiPhenix}. For 
$\Omega $ baryons, their mean transverse momentum is about $0.93$ and $0.83$
GeV/c for $b=0$ fm and $b=8$ fm, respectively, and are larger than those of $%
\phi $ mesons.

We note that hadron transverse momentum spectra obtained from the AMPT model
with string melting are generally softer than those measured in experiments 
\cite{chen05}. Since hadron spectra reflect those of partons in the parton
coalescence model, soft hadron spectra are due to soft parton spectra during
the partonic stage. As discussed in Ref. \cite{chen05}, the latter is a
result of the small current quark masses used in the parton cascade of the 
\textrm{AMPT} model with string melting, which make their transverse
momentum spectra less affected by radial collective flow than hadrons in the
default \textrm{AMPT} model. Since anisotropic flows are given by ratios of
hadron transverse momentum spectra, predictions of anisotropic flows as well
as the ratio of hadron transverse momentum spectra from present AMPT model
are expected to be more reliable \cite{Lin:2001zk,chen05}.

The ratio of midrapidity $\Omega $ baryons to $\phi $ mesons shown in the
lower panels of Fig. \ref{dNdPTratioPhiOmg} is seen to increase appreciably
from low $p_{T}$ to intermediate $p_{T}$ of about $2.5$ GeV/c. The
enhancement factor is about $7$ and $4$ for $b=0$ fm and $b=8$ fm,
respectively. This enhancement is a result of parton coalescence as in the
case of observed anomalously large anti-proton to pion ratio of about $0.8$
and $0.4$ at $p_{T}\gtrsim 2.5$ GeV/c in central and mid-peripheral
(centrality of about $30\%$) Au+Au collisions at $\sqrt{s_{NN}}=200$ GeV,
respectively \cite{ppiPhenix}. As explained in Refs. \cite{greco,hwa,fries},
in the coalescence model baryons with transverse momentum $p_{T}$ are mainly
formed from quarks with transverse momenta $\sim p_{T}/3$, while mesons with
same transverse momentum are mainly produced from partons with transverse
momenta $\sim p_{T}/2$. Since quark transverse momentum spectra decrease
with $p_{T}$, it is more favorable to produce high transverse momentum
baryons than mesons if hadrons are produced from the QGP through the
coalescence of quarks. However, the $\bar{\Omega}^{+}$/$\phi $ ratio is only
about $0.23$ and $0.17$ at $p_{T}\approx 2.5$ GeV/c for $b=0$ fm and $b=8$
fm, respectively, and is significantly smaller than the anti-proton to pion
ratio observed in experiments. For $b=0$ fm, our result with $R_{\phi }=0.65$
fm reproduces reasonably the measured $\Omega $/$\phi $ ratio as shown in
lower left panel of Fig. \ref{dNdPTratioPhiOmg}.

\begin{figure}[th]
\includegraphics[scale=0.95]{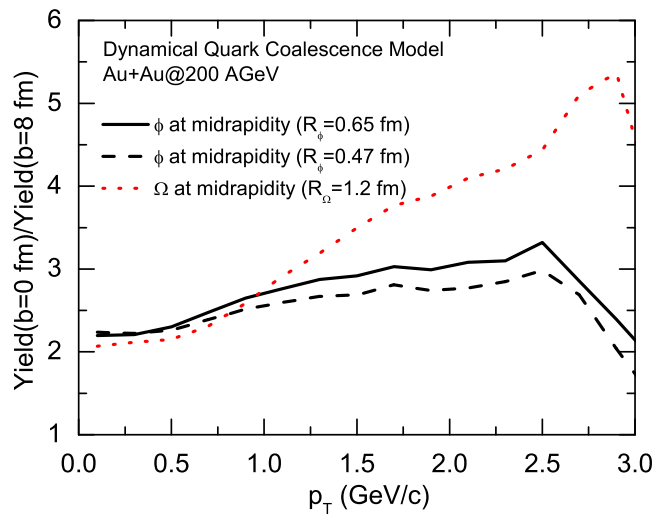}
\caption{{\protect\small (Color online) }{\protect\small Transverse momentum
dependence of the ratio of midrapidity $\protect\phi $ meson and $\Omega $
baryon yields in central (}$b=0${\protect\small \ fm) to that in
mid-peripheral (}$b=8${\protect\small \ fm) Au+Au collisions at }$\protect%
\sqrt{s_{NN}}=200${\protect\small \ GeV.}}
\label{RcpPTPhiOmg}
\end{figure}

In Fig. \ref{RcpPTPhiOmg}, we show the $p_{T}$ dependence of the ratio of
midrapidity $\phi $ meson or $\Omega $ baryon yield in central ($b=0$ fm) to
corresponding one in mid-peripheral ($b=8$ fm) collisions of Au+Au at $\sqrt{%
s_{NN}}=200$ GeV. It is seen that the $p_{T}$ dependence is very different
for $\phi $ mesons and $\Omega $ baryons with the latter having a smaller
value at low $p_{T}$ while a larger value at high $p_{T}$ than $\phi $
mesons. Our results thus indicate that nuclear suppression at intermediate
momenta is weaker for $\Omega $ baryons than $\phi $ mesons. This feature is
consistent with the measured $p_{T}$ dependence of so-called R$_{CP}$ (the
ratio of binary-collision-scaled $p_{T}$ spectra in central and peripheral
collisions) for $p+\bar{p}$ and pions \cite{ppiPhenix}.

\subsection{Anisotropic flows of $\protect\phi $ and $\Omega $}

\label{flow}

In heavy-ion collisions at RHIC, experimental data have indicated that there
exist not only strong elliptic flow but also a clear fourth-order
anisotropic flow $v_{4}$ for charged hadrons \cite{STAR03,STAR05}. An
interesting finding about $v_{4}$ is that it scales with the square of $%
v_{2} $, i.e., $v_{4}(p_{T})\sim v_{2}^{2}(p_{T})$. In the present
subsection, both elliptic and fourth-order anisotropic flows of $\phi $
mesons and $\Omega $ baryons are studied to see if they also show similar
features.

\begin{figure}[th]
\includegraphics[scale=0.85]{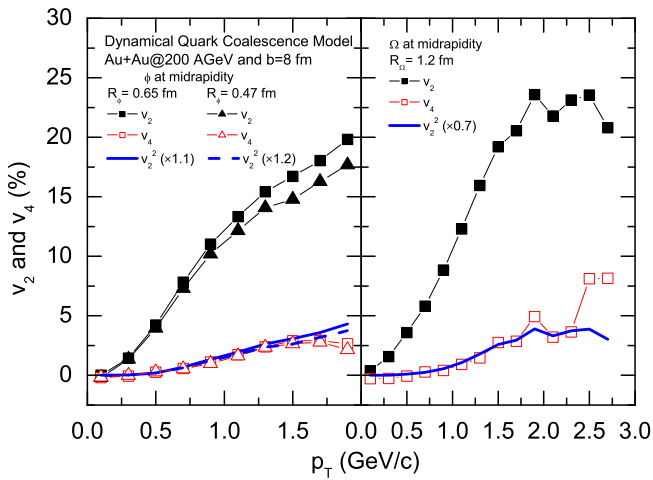}
\caption{{\protect\small (Color online) }{\protect\small Transverse momentum
dependence of anisotropic flows }$v_{2}${\protect\small \ and }$v_{4}$%
{\protect\small \ of mid-rapidity }$\protect\phi ${\protect\small \ mesons
(left panel) and $\Omega $ baryons (right panel) produced in Au+Au
collisions at }$\protect\sqrt{s_{NN}}=200${\protect\small \ GeV and }$b=8$%
{\protect\small \ fm. Solid and dashed lines in left panel are,
respectively, }$1.1v_{2}^{2}$ and $1.2v_{2}^{2}$ {\protect\small \ for } $%
\protect\phi ${\protect\small \ mesons, while solid line in right panel is }$%
0.7v_{2}^{2}${\protect\small \ \ for }$\Omega $ {\protect\small \ baryons.}}
\label{v24PTPhiOmg}
\end{figure}

In left panel of Fig. \ref{v24PTPhiOmg}, we show the $p_{T}$ dependence of
anisotropic flows $v_{2}$ and $v_{4}$ of mid-rapidity $\phi $ mesons
produced in Au+Au collisions at $\sqrt{s_{NN}}=200$ GeV and $b=8$ fm. Solid
and open squares are, respectively, the $\phi $ meson $v_{2}$ and $v_{4}$
for a $\phi $ meson radius $R_{\phi }=0.65$ fm, while solid and open
triangles are, respectively, those for $R_{\phi }=0.47$ fm. It is seen that
the scaling relation $v_{4}(p_{T})\sim v_{2}^{2}(p_{T})$ is satisfied in
both cases as shown by the solid ($1.1v_{2}^{2}$) and dashed ($1.2v_{2}^{2}$%
) lines in the same figure. It is interesting to see that the scaling
coefficient of $1.1$ or $1.2$ is the same as that extracted from the data
for charged hadrons \cite{STAR03,STAR05}. Similar results for $\Omega $
baryons using the radius $R_{\Omega }=1.2$ fm are shown in right panel of
Fig. \ref{v24PTPhiOmg}. The scaling relation $v_{4}(p_{T})\sim
v_{2}^{2}(p_{T})$ is again satisfied for $\Omega $ baryons but with a
smaller scaling coefficient than that for $\phi $ mesons as shown by the
solid line ($0.7v_{2}^{2}$) in the right panel.

Based on the naive momentum-space quark coalescence model that only allows
quarks with equal momentum to form hadrons \cite{molnar03}, this scaling
relation between hadron $v_{2}(p_{T})$ and $v_{4}(p_{t})$ has been
attributed to a similar scaling relation between those of quarks \cite%
{chen04,Kolb04}. Neglecting the small contribution of higher-order
anisotropic flows (higher than forth-order), this model gives for
mid-rapidity hadrons 
\begin{eqnarray}
\frac{v_{4,M}(2p_{T})}{v_{2,M}^{2}(2p_{T})} &\approx &\frac{1}{4}+\frac{1}{2}%
\frac{v_{4,q}(p_{T})}{v_{2,q}^{2}(p_{T})},\allowbreak  \notag
\label{v4Mscal} \\
\frac{v_{4,B}(3p_{T})}{v_{2,B}^{2}(3p_{T})} &\approx &\allowbreak \frac{1}{3}%
\left( 1+\frac{v_{4,q}(p_{T})}{v_{2,q}^{2}(p_{T})}\right) ,  \label{v4Bscal}
\end{eqnarray}%
where $v_{n,M}(p_{T})$, $v_{n,B}(p_{T})$ and $v_{n,q}(p_{T})$ denote,
respectively, the meson, baryon, and quark anisotropic flows. The hadron
anisotropic flows thus satisfy the scaling relation $v_{4}(p_{T})\sim
v_{2}^{2}(p_{T})$ if a similar scaling relation is satisfied by quark
anisotropic flows.

We have already shown in Fig. \ref{v24PTssBuuB} that the scaling relation $%
v_{4}(p_{T})\sim v_{2}^{2}(p_{T})$ is satisfied by mid-rapidity strange and
antistrange quarks ($s+\bar{s}$) with a scaling coefficient of about $0.85$,
i.e., $v_{4,q}(p_{T})/v_{2,q}^{2}(p_{T})\approx 0.85$. Eq. (\ref{v4Bscal})
then leads to a scaling coefficient of $v_{4,\phi }(p_{T})/v_{2,\phi
}^{2}(p_{T})\approx 0.68$ for $\phi $ mesons and $v_{4,\Omega
}(p_{T})/v_{2,\Omega }^{2}(p_{T})\approx 0.62$ for $\Omega $ baryons.
Comparing with the scaling coefficients of $1.1$ or $1.2$ for $\phi $ mesons
and $0.7$ for $\Omega $ baryons from the dynamical quark coalescence model,
the predicted value from the naive momentum-space quark coalescence model
for the $\phi$ meson scaling coefficient is significantly smaller. However,
the dynamical quark coalescence model does give a larger scaling coefficient
for $\phi $ mesons than for $\Omega $ baryons as expected from the naive
momentum-space coalescence model.

Another interesting and important finding in heavy-ion collisions at RHIC is
the valence quark number scaling of the elliptic flow of identified hadrons,
i.e., the elliptic flow per valence quark in a hadron is same at same
transverse momentum per valence quark, i.e., $%
v_{2,H}(p_{T}/n_{q})/n_{q}=v_{2,q}(p_{T})$ with $v_{2,H}$ and $n_{q}$
denoting, respectively, the hadron $v_{2}$ and the number of valence quarks
or anti-quarks in a hadron. In the naive momentum-space quark coalescence
model, this scaling has been shown to be satisfied if high-order anisotropic
flows are small \cite{molnar03,Kolb04}. The scaling is, however, broken in
more general quark coalescence models that take into account the quark
momentum distribution \cite{Greco:2004ex,Greco:2005jk,scott05} and higher
parton Fock states \cite{Muller:2005pv} in hadrons as well as the effect of
resonance decays \cite{Greco:2004ex}. It is thus of interest to see if the
elliptic flows of $\phi$ mesons and $\Omega $ baryons from present dynamical
quark coalescence model also satisfy the valence quark number scaling.

\begin{figure}[th]
\includegraphics[scale=0.9]{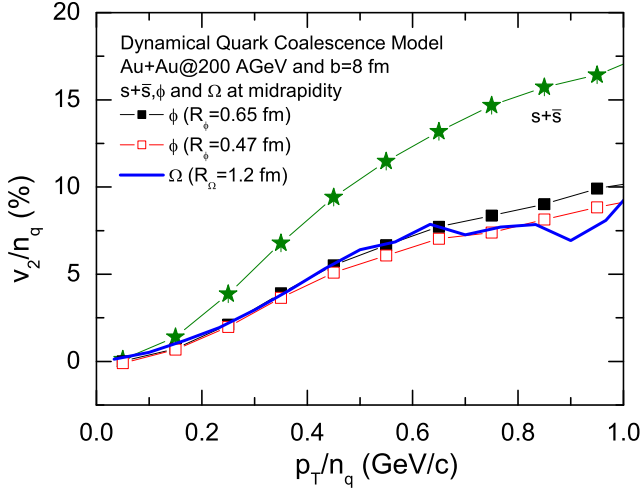}
\caption{{\protect\small (Color online) Valence quark number scaled elliptic
flow }$v_{2}/n_q${\protect\small \ as a function of scaled transverse
momentum }$p_T/n_q$ {\protect\small for mid-rapidity } $\protect\phi $%
{\protect\small \ mesons and }$\Omega ${\protect\small \ baryons produced in
Au+Au collisions at }$\protect\sqrt{s_{NN}}=200$ {\protect\small \ GeV and }$%
b=8${\protect\small \ fm with } $R_{\protect\phi }=0.65${\protect\small \ fm
(solid squares) and }$0.47$ {\protect\small \ fm (open squares) as well as }$%
R_{\Omega }=1.2$ {\protect\small \ fm (solid line). Results for mid-rapidity
strange and antistrange quarks (}$s+\bar{s}${\protect\small ) at freeze-out
are shown by solid stars.}}
\label{v2PTscaling}
\end{figure}

In Fig. \ref{v2PTscaling}, we show the valence quark number scaled elliptic
flow $v_{2}/n_{q}$ as a function of scaled transverse momentum $p_{T}/n_{q}$
for mid-rapidity $\phi $ mesons and $\Omega $ baryons produced in Au+Au
collisions at $\sqrt{s_{NN}}=200$ GeV and $b=8$ fm with $R_{\phi }=0.47$ or $%
0.65$ fm and $R_{\Omega }=1.2$ fm. For comparison, we also include the
elliptic flow of mid-rapidity strange and antistrange quarks ($s+\bar{s}$)
at freeze-out. It is seen that the elliptic flows of both mid-rapidity $\phi 
$ mesons and $\Omega $ baryons from the dynamical quark coalescence model
satisfy the valence quark number scaling. The valence quark number scaled
elliptic flows of $\phi $ mesons and $\Omega $ baryons are, however,
significantly smaller than that of coalescing strange and antistrange
quarks. This feature is different from the prediction of the naive
momentum-space quark coalescence model where the valence quark number scaled 
$v_{2}$ of hadrons is equal to the $v_{2}$ of coalescing quarks. Our results
therefore indicate that both the dynamical phase-space information of quarks
and the quark phase-space distribution inside hadrons play important roles
in hadron anisotropic flows, and cautions are needed in interpreting the
elliptic flow of partons from that of hadrons using the naive momentum-space
parton coalescence model. We note that our results are consistent with
previous studies on $v_{2}$ of protons and pions based on the parton
phase-space information from Molnar's parton cascade (MPC) model \cite%
{molnar05}.

\section{Hadron size dependence of $\protect\phi $ and $\Omega $ yields and
anisotropic flows}

\label{size}

In above calculations, we have used $\phi $ meson and $\Omega $ baryon size
parameters that are fixed from fitting measured yields in experiments.
Ideally, we would like to use the empirically and/or theoretically
determined root-mean-square radii of $\phi $ meson and $\Omega $ baryon in
the dynamical quark coalescence model. In the absence of such information,
it is of interest to study how above results depend on the sizes of $\phi $
meson and $\Omega $ baryon.

\subsection{Hadron size dependence of $\protect\phi $ and $\Omega $ yields}

\begin{figure}[th]
\includegraphics[scale=1.15]{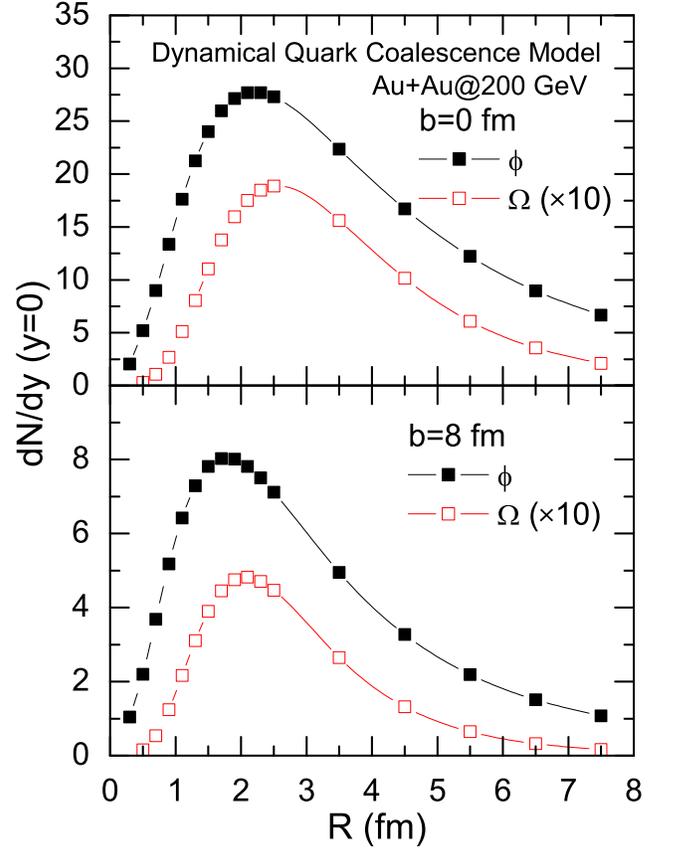}
\caption{{\protect\small (Color online) Hadron size (root-mean-square
radius) dependence of rapidity density $dN/dy$ of mid-rapidity }$\protect%
\phi ${\protect\small \ mesons (solid squares) and }$\Omega ${\protect\small %
\ (}$\Omega ^{-}+\bar{\Omega}^{+}${\protect\small ) baryons (open squares)
produced in Au+Au collisions at }$\protect\sqrt{s_{NN}}=200${\protect\small %
\ GeV with }$b=0${\protect\small \ fm (upper panel) and }$b=8 $%
{\protect\small \ fm (lower panel).}}
\label{dNdYPhiOmgR}
\end{figure}

We show in Fig. \ref{dNdYPhiOmgR} the rapidity density $dN/dy$ of
mid-rapidity $\phi $ mesons and $\Omega $ baryons produced in Au+Au
collisions at $\sqrt{s_{NN}}=200$ GeV with $b=0$ fm (upper panel) and $b=8$
fm (lower panel) as functions of their root-mean-square radii. It is seen
that both $\phi$ meson and $\Omega$ baryon yields are sensitive to their
sizes. For physically reasonable radii, both yields increase with increasing
radii but decrease eventually as the radii become unrealistically large. For
collisions at impact parameter $b=0$ fm, the yield exhibits a maximum at $%
R_{\phi }=2.2$ fm for $\phi $ mesons and at $R_{\Omega }=2.5$ fm for $\Omega$
baryons. A similar feature is observed in the case of $b=8$ fm but the
maximum is at $R_{\phi }=1.75$ fm for $\phi $ mesons and at $R_{\Omega }=2$
fm for $\Omega $ baryons.

\subsection{Hadron size dependence of $\protect\phi $ and $\Omega $ elliptic
flows and the valence quark number scaling}

\begin{figure}[th]
\includegraphics[scale=0.85]{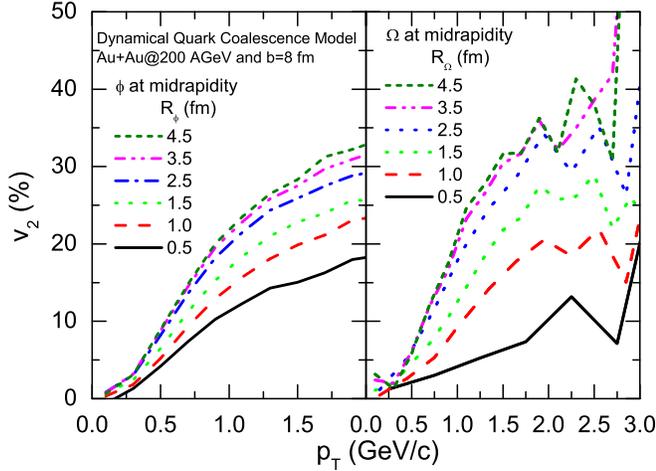}
\caption{{\protect\small (Color online) }{\protect\small Transverse momentum
dependence of the }$v_{2}${\protect\small \ of mid-rapidity }$\protect\phi $%
{\protect\small \ mesons (left panel) and $\Omega $ baryons (right panel) in
Au+Au collisions at }$\protect\sqrt{s_{NN}}=200$ {\protect\small \ GeV and }$%
b=8${\protect\small \ fm with different values of root-mean-square radii $R_{%
\protect\phi }$ and $R_{\Omega }$}.}
\label{v2PTPhiOmgR}
\end{figure}

\begin{figure}[th]
\includegraphics[scale=0.9]{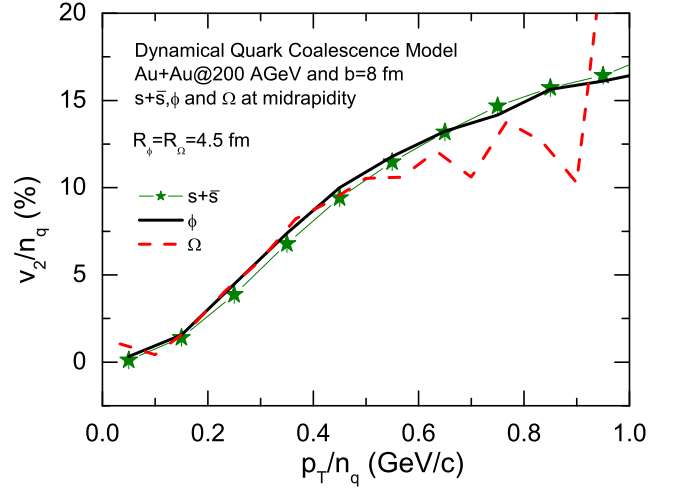}
\caption{{\protect\small (Color online) Same as Fig. \protect\ref%
{v2PTscaling} but with }$R_{\protect\phi }=${\protect\small \ }$R_{\Omega
}=4.5${\protect\small \ fm.} }
\label{v2PTscalingR45}
\end{figure}

We first show in left panel of Fig. \ref{v2PTPhiOmgR} the $p_{T}$-dependence
of the $v_{2}$ of mid-rapidity $\phi $ mesons in Au+Au collisions at $\sqrt{%
s_{NN}}=200$ GeV and $b=8$ fm for different values of root-mean-square
radius $R_{\phi }$. It is clearly seen that the $v_{2}$ of $\phi $ mesons
depends strongly on the $\phi $ meson size. It becomes larger with
increasing $R_{\phi }$ but changes little when $R_{\phi }>4.5$ fm.
Similarly, we show in right panel of Fig. \ref{v2PTPhiOmgR} the $p_{T}$%
-dependence of the $v_{2}$ of mid-rapidity $\Omega $ baryons produced in the
same collision with different values of $R_{\Omega }$. Similar to that of $%
\phi $ mesons, the $v_{2}$ of $\Omega $ baryons also increases with
increasing size of $\Omega $ baryon and saturates when $R_{\Omega }>4.5$ fm.
The saturation of $\phi $ meson and $\Omega $ baryon $v_{2}$ at large sizes
observed in Fig. \ref{v2PTPhiOmgR} is essentially due to the fact that the
dynamical quark coalescence model approaches the naive momentum-space quark
coalescence model, i.e, only partons with same momentum can coalesce into
hadrons, when the hadron size is large enough. This can be easily seen from
Eqs. (\ref{WignerPhi}) and (\ref{WignerOMG}) as the quark Wigner phase-space
function becomes essentially delta functions in the relative momenta of
coalescing partons when hadron sizes (i.e., the parameters $\sigma _{\phi }$%
\ and $\sigma _{\Omega }$) are very large. As a result, we expect that the
valence quark number scaled elliptic flows of $\phi $ mesons and $\Omega $
baryons should approach the $v_{2}$ of coalescing strange and antistrange
quarks as their sizes increase. This is shown in Fig. \ref{v2PTscalingR45}
where it is seen that the $p_{T}/n_{q}$-dependence of $v_{2}/n_{q}$ for
mid-rapidity $\phi $ mesons and $\Omega $ baryons produced in Au+Au
collisions at $\sqrt{s_{NN}}=200$ GeV and $b=8$ fm with $R_{\phi }=R_{\Omega
}=4.5$ fm is indeed similar to the elliptic flow of mid-rapidity strange and
antistrange quarks ($s+\bar{s}$) at freeze out.

\subsection{Fourth-order flows of $\protect\phi $ and $\Omega $}

\begin{figure}[th]
\includegraphics[scale=0.85]{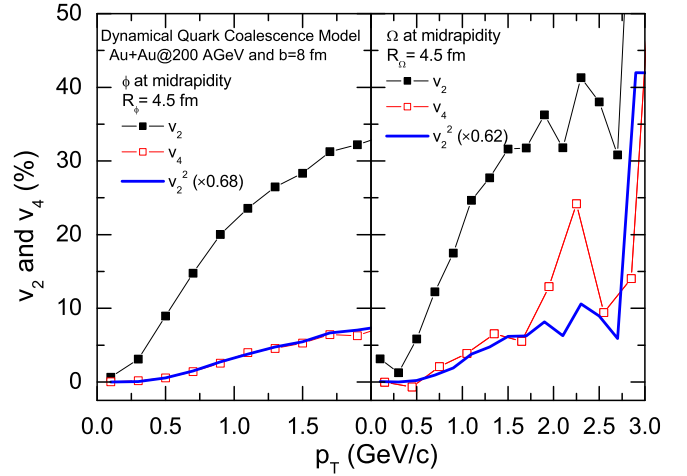}
\caption{{\protect\small (Color online) Same as Fig. \protect\ref%
{v24PTPhiOmg} but with }$R_{\protect\phi }=R_{\Omega }=4.5${\protect\small \
fm.}}
\label{v24PTR45}
\end{figure}

For the scaling relation $v_{4}(p_{T})\sim v_{2}^{2}(p_{T})$ among hadron
anisotropic flows, as discussed above and in Refs. \cite{chen04,Kolb04}, the
dynamical quark coalescence model leads to a deviation of the scaling
coefficient from that of the naive momentum-space quark coalescence model as
a result of finite hadron sizes and nonzero parton relative momenta inside
hadrons. The deviation is expected to disappear for large enough hadron
sizes when the dynamical quark coalescence model approaches the naive
momentum-space quark coalescence model. This is demonstrated in Fig. \ref%
{v24PTR45} where we show the $p_{T}$-dependence of the anisotropic flows $%
v_{2}$ (solid squares) and $v_{4}$ (open squares) of mid-rapidity $\phi $
mesons (left panel) and $\Omega $ baryons (right panel) produced in Au+Au
collisions at $\sqrt{s_{NN}}=200$ GeV and $b=8$ fm with large hadron sizes,
i.e., $R_{\phi }=R_{\Omega }=4.5$ fm. Also shown by solid lines are $%
0.68v_{2}^{2}$ for $\phi$ mesons in left panel and $0.62v_{2}^{2}$ for $%
\Omega$ baryons in right panel. It is seen that for such large hadron sizes
the scaling coefficients indeed approach the values expected from the naive
momentum-space quark coalescence model, i.e., about $0.68$ for $\phi $
mesons and $0.62$ for $\Omega $ baryons. We note that a larger hadron size
generally gives a smaller scaling coefficient in the dynamical quark
coalescence model.

\section{Quark mass dependence of $\protect\phi $ and $\Omega $ transverse
momentum spectra and anisotropic flows}

\label{mass}

Above results were obtained with the current strange quark mass of $199$ 
\textrm{MeV}, which is the default value in the AMPT model with string
melting. Usually, the constituent quark mass (about $500$ \textrm{MeV} for
the strange quark) is used in the quark coalescence model \cite%
{greco,hwa,fries,molnar03}, so the binding energy effect can be neglected.
In the following, we study the dependence of $\phi $ meson and $\Omega $
baryon production and their anisotropic flows on the strange quark mass. In
particular, we use the constituent strange quark mass of $500$ \textrm{MeV}
but same strange quark space-time coordinates and momenta at freeze out as
before. The quark mass effect thus shows up through the Lorentz
transformation that changes the strange quark space-time coordinates and
momenta in the rest frame of the $\phi $ meson and $\Omega $ baryon and
therefore affects their overlap with the quark Wigner phase-space functions
inside hadrons.

\subsection{Yields and transverse momentum spectra of $\protect\phi$ and $%
\Omega$}

Using a strange quark mass of $500$ \textrm{MeV}, we find that yields of
both $\phi $ mesons and $\Omega $ baryons are enhanced significantly
compared to previous results with a strange quark mass of 199 MeV. For Au+Au
collisions at $\sqrt{s_{NN}}=200$ GeV and $b=0$ ($8$) fm, the $dN/dy$ at
mid-rapidity for $\phi $ mesons with $R_{\phi }=0.65$ fm changes from $8.0$ (%
$3.3$) to $12.2$ ($4.8$) while that with $R_{\phi }=0.47$ fm changes from $%
4.7$ ($2.0$) to $7.3$ ($3.0$) when the strange quark mass changes from $199$
MeV to $500$ \textrm{MeV}. To keep the $dN/dy$ of $\phi $ mesons at
mid-rapidity unchanged requires a reduction of the $\phi $ meson size from $%
R_{\phi }=0.65$ fm to 0.5 fm or from $R_{\phi }=0.47$ fm to $0.35$ fm when
the strange quark mass is increased to $500$ \textrm{MeV}. For $\Omega $
baryons, the value of $dN/dy$ at mid-rapidity for $b=0$ ($8$) fm with $%
R_{\Omega }=$ $1.2$ fm changes from $0.65$ ($0.26$) to $1.36$ ($0.49$) when
the strange quark mass changes from $199$ MeV to $500$ \textrm{MeV}. A
somewhat smaller $\Omega $ baryon size of $R_{\Omega }=0.9$ fm is then
needed to keep $dN/dy$ of $\Omega $ baryons at mid-rapidity unchanged when
the strange quark mass is increased to $500$ \textrm{MeV}.

\begin{figure}[th]
\includegraphics[scale=0.95]{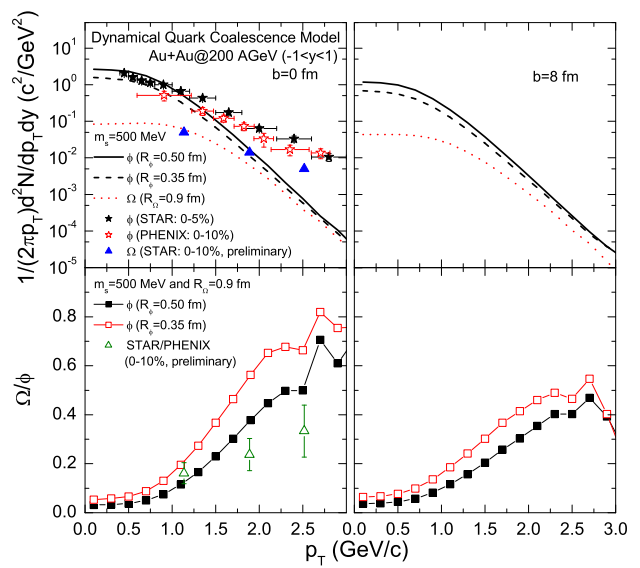}
\caption{{\protect\small (Color online) Same as Fig. \protect\ref%
{dNdPTratioPhiOmg} but with }$m_{s}=500${\protect\small \ MeV and }$R_{%
\protect\phi }=0.5${\protect\small \ fm or }$0.35${\protect\small \ fm and }$%
R_{\Omega }=0.9${\protect\small \ fm.} }
\label{dNdPTratioPhiOmgMS5h}
\end{figure}

We also find that the larger strange quark mass of $500$ \textrm{MeV} leads
to a slightly stiffer transverse momentum spectra for $\phi $ mesons and $%
\Omega $ baryons. For $R_{\phi }=0.65$ ($0.47$) fm, the mean transverse
momentum $\langle p_{T}\rangle $ for the $\phi $ meson changes from about $%
0.72$ ($0.73$) and $0.68$ ($0.70$) GeV/c for $b=0$ fm and $b=8$ fm,
respectively, to about $0.74$ ($0.76$) and $0.69$ ($0.72$) GeV/c for $b=0$
fm and $b=8$ fm, respectively, when the strange quark mass changes from $199$
MeV to $500$ \textrm{MeV}. For $\Omega $ baryons, the value of $\langle
p_{T}\rangle $ with $R_{\Omega }=$ $1.2$ fm changes from about $0.93$ and $%
0.83$ GeV/c for $b=0$ fm and $b=8$ fm, respectively, to about $0.96$ and $%
0.85$ GeV/c for $b=0$ fm and $b=8$ fm, respectively, when the strange quark
mass changes from $199$ MeV to $500$ \textrm{MeV}.

With the new parameters $m_{s}=500$ MeV, $R_{\phi }=0.5$ fm or $0.35$ fm,
and $R_{\Omega }=0.9$ fm fixed from measured yields of $\phi $ mesons and $%
\Omega $ baryons, the transverse momentum spectra of midrapidity $\phi $
mesons and $\Omega $ baryons and their ratio in the same collision are shown
in Fig. \ref{dNdPTratioPhiOmgMS5h}. As in Fig. \ref{dNdPTratioPhiOmg},
corresponding experimental data are also included in left panels of Fig. \ref%
{dNdPTratioPhiOmgMS5h}. For $R_{\phi }=0.5$ ($0.35$) fm and $m_{s}=500$ MeV,
the mean $\phi $ meson transverse momentum calculated from the $\phi $ meson
transverse momentum spectrum is about $0.76 $ ($0.77$) and $0.72$ ($0.74$)
GeV/c for $b=0$ fm and $b=8$ fm, respectively, which are still somewhat
smaller than the experimental value of about $0.85$-$1.1$ GeV/c \cite%
{phiStar,phiPhenix}. For $\Omega $ baryons with $R_{\Omega }=$ $0.9$ fm,
their mean transverse momentum is about $1.05$ and $0.94$ GeV/c for $b=0$ fm
and $b=8$ fm, respectively.

The $\Omega $/$\phi $ ratio obtained from above new parameters is shown in
lower panels of Fig. \ref{dNdPTratioPhiOmgMS5h}. Compared with the case of $%
m_{s}=199$ MeV, the $\Omega /\phi $ ratio increases even more from low $%
p_{T} $ to intermediate $p_{T}$ of about $2.5$ GeV/c, leading to a larger
enhancement factor about $14$ and $9$ for $b=0$ fm and $b=8$ fm,
respectively. The $\bar{\Omega}^{+}$/$\phi $ at $p_{T}\approx 2.5$ GeV/c is
now about $0.30$ and $0.22$ for $b=0$ fm and $b=8$ fm, respectively, which
are still significantly smaller than that of anti-protons to pions observed
in experiments but remain comparable to the measured $\Omega /\phi $ ratio
for the larger $\phi $ meson size of $R_{\phi }=0.5$ fm.

\subsection{Anisotropic flows of $\protect\phi$ and $\Omega$}

\begin{figure}[th]
\includegraphics[scale=0.85]{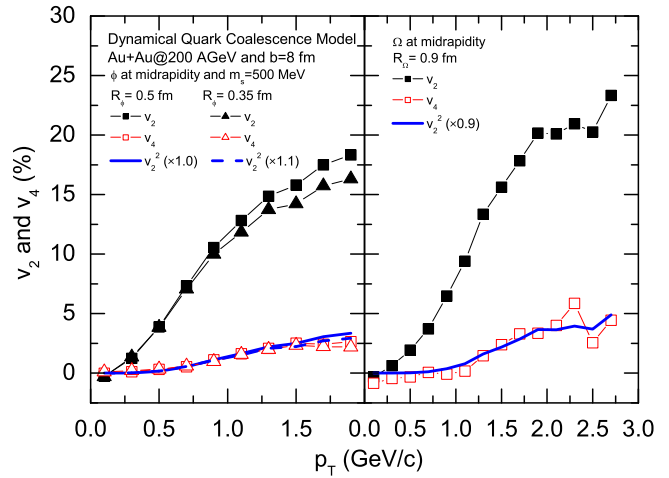}
\caption{{\protect\small (Color online) Same as Fig. \protect\ref%
{v24PTPhiOmg} but with }$m_{s}=500${\protect\small \ MeV and }$R_{\protect%
\phi }=0.5${\protect\small \ fm or }$0.35${\protect\small \ fm and }$%
R_{\Omega }=0.9${\protect\small \ fm.}}
\label{v24PTPhiOmgMS5h}
\end{figure}

Using the fitted hadron size parameters for $m_{s}=500$ MeV, i.e., $R_{\phi
}=0.5$ fm or $0.35$ fm and $R_{\Omega }=0.9$ fm, we show in left panel of
Fig. \ref{v24PTPhiOmgMS5h} the $p_{T}$ dependence of the anisotropic flows $%
v_{2}$ (solid squares for $R_{\phi }=0.5$ fm and triangles for $R_{\phi
}=0.35$ fm) and $v_{4}$ (open squares for $R_{\phi }=0.5$ fm and triangles
for $R_{\phi }=0.35$ fm) of mid-rapidity $\phi $ mesons produced in Au+Au
collisions at $\sqrt{s_{NN}}=200$ GeV and $b=8$ fm. Also shown in left panel
of Fig. \ref{v24PTPhiOmgMS5h} are $1.0v_{2}^{2}$ (solid line for $R_{\phi
}=0.5$ fm) and $1.1v_{2}^{2}$ (dashed line for $R_{\phi }=0.35$ fm). Similar
results for $\Omega $ baryons are shown in right panel of Fig. \ref%
{v24PTPhiOmgMS5h}. Compared to those shown in Fig. \ref{v24PTPhiOmg} with a
strange quark mass of $199$ MeV, the hadron elliptic flow is seen to depend
only weakly on the strange quark mass, with the larger strange quark mass of 
$m_{s}=500$ MeV giving a slightly smaller $v_{2}$ at lower $p_{T}$ while a
slightly larger $v_{2}$ at higher $p_{T}$. Also, the anisotropic flows of $%
\phi $ mesons and $\Omega $ baryons from the dynamical quark coalescence
model with the larger strange quark mass of $m_{s}=500$ MeV still satisfy
the scaling relation $v_{4}(p_{T})\sim v_{2}^{2}(p_{T})$ with similar
scaling coefficients as those for the strange quark mass of 199 MeV.

\begin{figure}[th]
\includegraphics[scale=0.9]{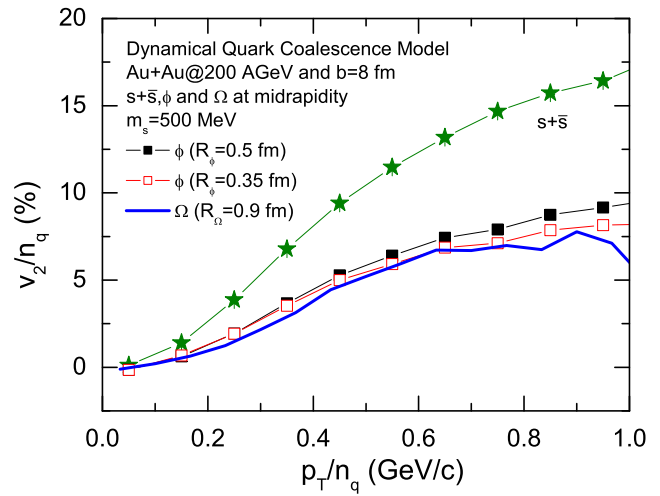}
\caption{{\protect\small (Color online) Same as Fig. \protect\ref%
{v2PTscaling} but with }$m_{s}=500${\protect\small \ MeV and }$R_{\protect%
\phi }=0.5${\protect\small \ fm and }$0.35${\protect\small \ fm as well as }$%
R_{\Omega }=0.9${\protect\small \ fm.}}
\label{v2PTscalingMS5h}
\end{figure}

The $p_{T}/n_{q}$-dependence of $v_{2}$ per valence quarks or anti-quarks
for mid-rapidity $\phi $ mesons and $\Omega $ baryons produced in Au+Au
collisions at $\sqrt{s_{NN}}=200$ GeV and $b=8$ fm, obtained with the large
strange quark mass of $500$ MeV, $R_{\phi }=0.5$ fm or $0.35$ fm and $%
R_{\Omega }=0.9$ fm, is shown in Fig. \ref{v2PTscalingMS5h} together with
the elliptic flow of mid-rapidity strange and antistrange quarks ($s+\bar{s}$%
) at freeze-out. It is seen that the $v_{2}$ of mid-rapidity $\phi $ mesons
and $\Omega $ baryons still satisfies approximately the valence number
scaling in the dynamical quark coalescence model. However, the valence quark
number scaled $v_{2}$ remains significantly smaller than the $v_{2}$ of
coalescing strange and antistrange quarks. 

\section{Summary}

\label{summary}

Based on the parton phase-space information obtained from a multi-phase
transport model within the string melting scenario, we have studied the
production of $\phi $ mesons and $\Omega $ baryons and their anisotropic
flows in Au+Au collisions at RHIC using a dynamical quark coalescence model,
which requires information on the radii of $\phi $ meson and $\Omega $
baryon. Fixing their radii by fitting measured yields of $\phi $ mesons and $%
\Omega $ baryons at midrapidity in central Au+Au collisions at $\sqrt{s_{NN}}%
=200$ GeV, we have evaluated their transverse momentum spectra in the same
collision at impact parameter $b=0$ and $8$ fm and also their anisotropic
flows in the same collision at impact parameter $b=8$ fm. Comparing with its
value at low transverse momenta, we have found that the ratio of the yield
of $\Omega $ baryons to that of $\phi $ mesons is enhanced significantly at
intermediate transverse momenta as observed in experiments.

We have further found that the elliptic flows of $\phi $ mesons and $\Omega $
baryons follow approximately the valence quark number scaling. The valence
quark number scaled elliptic flows of $\phi $ mesons and $\Omega $ baryons
deviate, however, strongly from the underlying $v_{2}$ of strange and
antistrange quarks. Moreover, we have also studied the forth-order
anisotropic flow $v_{4}$ and found that the scaling relation of $%
v_{4}(p_{T})\sim v_{2}^{2}(p_{T})$ observed experimentally for charged
hadrons is satisfied by $\phi $ mesons and $\Omega $ baryons as well. It
will be very interesting to compare these predictions with experimental data
that are being analyzed.

In addition, we have studied the dependence of above results on the radii of 
$\phi $ meson and $\Omega $ baryon as well as the strange quark mass. Both
the yields and anisotropic flows of $\phi $ mesons and $\Omega $ baryons are
found to be sensitive to their radii. For sufficient large radii, the
valence quark number scaled elliptic flows of $\phi $ mesons and $\Omega $
baryons approach the $v_{2}$ of strange and antistrange quarks as in the
naive momentum-space coalescence model. Also the scaling coefficient $%
v_{4}(p_{T})/v_{2}^{2}(p_{T})$ is sensitive to hadron sizes. Although the
strange quark mass was found to affect significantly the yields of $\phi $
mesons and $\Omega $ baryons, it does not change much their anisotropic
flows. Our results thus suggest that in using the quark coalescence model to
extract the parton dynamics in relativistic heavy-ion collisions from hadron
observables, it is important to take into account the quark structure of
hadrons.

Although results from present study reproduce reasonably the observed
transverse momentum dependence of the $\Omega /\phi $ ratio, the transverse
momentum spectra of $\phi $ mesons and $\Omega $ baryons are too soft
compared with measured ones. This has been attributed to the small current
quark masses used in the AMPT model. Recent studies have shown that the
equation of state of the QGP from lattice QCD calculations can only be
reproduced by partons with large masses \cite{Levai:1997yx,Peshier:2005pp}.
It is thus of great interest to improve the AMPT model by using massive
partons and to study how the resulting parton dynamics affects the results
obtained in present study. Although we have shown in present study that the
anisotropic flows of $\phi $ mesons and $\Omega $ baryons are not sensitive
to the strange quark mass used in the dynamical coalescence model if same
strange quark distribution is used, using massive partons in the AMPT model
not only is expected to affect the transverse momentum spectra of partons
but also may influence their anisotropic flow and thus that of produced
hadrons. Also, it is important to include in future studies the contribution
from high momentum jets \cite{Greco:2004yc} as they are expected to affect
the production of $\phi $ mesons and $\Omega $ baryons as well as their
anisotropic flows at intermediate and high transverse momenta.

\begin{acknowledgments}
We thank Professor Zong-Ye Zhang for helpful communications on the radius of
the $\Omega $ baryon in the quark model. This work was supported in part by
the National Natural Science Foundation of China under Grant Nos. 10105008
and 10575071 and by MOE of China under project NCET-05-0392 (LWC) as well as
by the US National Science Foundation under Grant No. PHY-0457265 and the
Welch Foundation under Grant No. A-1358 (CMK).
\end{acknowledgments}

\end{document}